\documentclass{PoS}

\title{Dark Matter}

\ShortTitle{Dark Matter}

\author{\speaker{Annika H. G. Peter}%
         \thanks{I would like to thank the organizers and staff at UT Austin for inviting me to and running an excellent symposium.  I would like to acknowlege my collaborators James Bullock, Manoj Kaplinghat, Jonathan Loranger, Leonidas Moustakas, Miguel Rocha, and Kris Sigurdson for conversations that were useful for crafting the talk, as well as providing some of the plots that were in the talk.  I am supported by a Gary McCue Fellowship through the Center for Cosmology at UC Irvine and NASA Grant No. NNX09AD09G.}\\
        Department of Physics and Astronomy, University of California, Irvine, CA 92697-4575\\
        E-mail: \email{annika.peter@uci.edu}}

\abstract{From astronomical observations, we know that dark matter exists, makes up 23\% of the mass budget of the Universe, clusters strongly to form the load-bearing frame of structure for galaxy formation, and hardly interacts with ordinary matter except gravitationally.  However, this information is not enough to identify the particle specie(s) that make up dark matter.  As such, the problem of determining the identity of dark matter has largely shifted to the fields of astroparticle and particle physics.  In this talk, I will review the current status of the search for the nature of dark matter.  I will provide an introduction to possible particle candidates for dark matter and highlight recent experimental astroparticle- and particle-physics results that constrain the properties of those candidates.  Given the absence of detections in those experiments, I will advocate a return of the problem of dark-matter identification to astronomy, and show what kinds of theoretical and observational work might be used to pin down the nature of dark matter once and for all.  This talk is intended for a broad astronomy audience.
}

\FullConference{Frank N. Bash Symposium 2011: New Horizons in Astronomy\\
		 October 9-11, 2011 \\
		 Austin, Texas, USA}

\begin{document}

\section{Introduction}\label{sec:intro}
Dark matter is the dominant gravitationally attractive component in the Universe, but we do not know what it consists of.  All the evidence for the existence of dark matter and constraints on its nature come from astronomy.  This is what we know so far:

\begin{figure}
\centering \includegraphics[width=0.4\textwidth,angle=270]{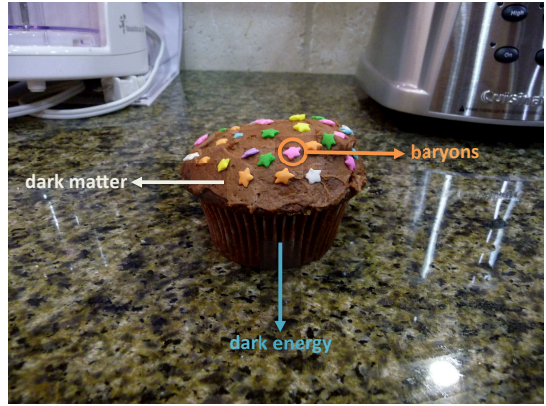}
\caption{\label{fig:cupcake}Contents of the Universe, as illustrated by a chocolate cupcake. Recipe available upon request.}
\end{figure}

\textbf{Abundance:} We may infer the content of the Universe from observations of the cosmic microwave background and of large-scale structure \cite{lss,cmb,cmbtutorial}.  The relative abundances of the major components of the Universe are illustrated by the chocolate cupcake in Fig. \ref{fig:cupcake}.  Baryons lightly sprinkle the Universe, as they constitute only about 4\% of the total mass-energy density.  Dark energy makes up the bulk of the Universe at the present epoch, clocking in at $\sim 73\%$, just as the cake dominates the cupcake.  Dark matter comprises $\sim23\%$ of the Universe.  Just as the chocolate frosting glues the sprinkles together on the cupcake, dark matter holds baryons together to form galaxies, galaxy groups, and galaxy clusters.

\textbf{A few things it cannot be:} Dark matter cannot consist of baryons.  There are two lines of evidence for this.  First, if baryons made up all the dark matter, the cosmic microwave background and cosmic web of structure would look radically different.  Second, the abundance of light elements created during big-bang nucleosynthesis depends strongly on the baryon density (more precisely, on the baryon-to-photon ratio) of the Universe (see \cite{steigman2007} and references therein).  Observed abundances of deuterium and $^4$He constrain give similar constraints on the baryon density in the Universe as those coming from cosmic microwave background observations. These lines of evidence imply that a once-popular class of baryonic dark-matter candidate, the Massive Compact Halo Object (MaCHO) class (e.g., brown dwarfs, stellar remnants) is cosmologically insignificant.

Dark matter cannot consist of light (sub-keV-mass) particles unless they were created via a phase transition in the early Universe (like QCD axions \cite{axions}).  This is because light particles are relativistic at early times, and thus fly out of small-scale density perturbations.  If particles were created thermally or via neutrino oscillations, the speed of the particles, and hence the distance they stream out of density perturbations, should be correlated with their mass.  Thus, one may map the smallest distance scale on which one sees clumpy structure to set a lower limit on the dark-matter particle mass (low mass == high speed == large distance traveled == scale on which density perturbations are washed out).  Current measurements of the Lyman-$\alpha$ forest, a probe of small-scale structures at $z\sim 3$, constrain the particle mass to be $m \gtrsim 2$ keV \cite{lya_darkmatter}.

\textbf{Electromagnetic neutrality:} There are strong constraints on the electromagnetics of dark matter \cite{em}.  If dark matter had either a small charge or a small electric or magnetic dipole moment, it would couple to the photon-baryon fluid before recombination, thus altering the sub-degree-scale features of the cosmic microwave background as well as the matter power spectrum.

\textbf{Self-interaction constraints:} Dark matter is part of a new sector of physics.  We may generically expect that dark-matter particles might interact with themselves or other new particles, mediated by new, dark gauge bosons.  Even if the particles in the dark sector have no coupling to the Standard Model (i.e., the particles and forces we know of), such interactions will affect the structures of dark-matter halos, since dark-matter particles may transfer energy and angular momentum in the scatters \cite{sidmth}.  For hard-sphere elastic scattering, the constraints are at the level of cross section per unit particle mass $\sigma/m \lesssim 1\hbox{ cm}^2/\hbox{g}$ from observations of the structure of galaxy clusters \cite{sidmobs}.

\textbf{Clumping on small scales:} There is evidence for virialized structures of dark matter down to scales of $\sim 10^7 -10^9 M_\odot$ halos.  Sophisticated modeling of the lens galaxy SDSSJ0946+1006 (a rare double Einstein ring system) indicates that it contains a dark-matter clump of mass $\sim 3\times 10^{9} M_\odot$ \cite{vegetti}.  The Milky Way dwarf galaxies are both the most dark-matter-dominated structures known (see \cite{mwdwarfs} and references therein).  Within their half-light radii of $\sim 30-800$ pc, they contain $\sim 10^6-10^8M_\odot$ of dark matter, with a mass-to-light ratio $\Upsilon_{1/2} \sim 10-4000\Upsilon_\odot$.  These galaxies are hosted by halos that were of order $10^9-10^{10}M_\odot$ before accretion onto the Milky Way.

\vskip 0.2cm

In this admittedly biased walk through the state of the dark-matter identification landscape, I start in Sec. \ref{sec:particle} by introducing popular dark-matter candidates.  In Sec. \ref{sec:astroparticle}, I describe dark-matter searches that rely on dark matter's non-gravitational interactions with the Standard Model.  In Sec. \ref{sec:astro}, I describe how to further exploit astronomical observations to uncover dark-matter physics.  In Sec. \ref{sec:conclusion}, I show how a synthesis of these approaches is needed to characterize dark matter.  

\section{The particle zoo}\label{sec:particle}

The only major non-particle candidate for dark matter is the primordial black hole, which would have collapsed directly from highly overdense regions of the early Universe, the existence of which requires funky physics \cite{pbh}.  At the risk of offending some of my colleagues, I claim that the only \emph{really plausible} dark-matter candidates are new particles.

I sometimes joke, there must be at least one candidate per particle model builder.  Nevertheless, there is a hierarchy among the particle candidates.  The top tier of candidates are called ``natural'' dark-matter candidates.  I call them the ``buy one, get one free'' candidates because we get these candidates ``for free'' from theories that solve other deep problems in physics.  Here are the most popular ``buy one, get one free'' candidates or classes of candidate:

\textbf{Weakly-interacting massive particles (WIMPs):} This class of candidate, or at least its delightful moniker, was originally introduced by Steigman \& Turner \cite{steigman1985}.  This key features of this particle class are exactly as described: interactions around or near typical weak-force interactions (the fine-structure constant $\alpha$ near the weak-scale coupling $\sim 10^{-2}$), particle masses near the weak scale ($m\sim 100$ GeV in particle-physics units \cite{astronomerunit}, similar to the mass of a silver atom).  

This candidate class has the additional feature that it may ``naturally'' make up all the dark matter, thus making it more ``Black Friday sale'' dark matter than the ``buy one, get one free'' candidate.  This feature of WIMPs is called the ``WIMP miracle''.  The origin of the WIMP miracle is this.  If WIMPs are in a thermal bath in the early Universe with other particles, having been born out of decays of the inflaton or something of the like, then we can solve Boltzmann equations to find that WIMPs ``freeze out'' (i.e., stop being created/destroyed through annihilations with other particles) at a comoving density that is inversely proportional to the WIMP annihilation cross section $\sigma_{ann}$.  Unless decays are important, this comoving number density is fixed for all future time.  By dimensional analysis (recalling that mass is inversely proportional to the length scale in particle-physics units), the annihilation cross section should be $\sigma_{ann} \propto \alpha^2/m^2$.  If you put this dimensional-analysis cross section into early-Universe Boltzmann equations, the comoving number density of WIMPs matches the number density inferred from cosmological observations \cite{lss,cmb}.  


Candidates in the WIMP class include the supersymmetric neutralino (the lowest-mass eigenstate of the supersymmetric partners of neutral Standard Model gauge bosons) and the Kaluza-Klein photon \cite{griest}.  Both of these candidates emerge out of theories to introduce new physics at the electroweak breaking scale (the minimal supersymmetric standard model [MSSM] and universal extra dimensions [UED]), and possibly to explain why that scale is so much lower than the Planck scale.  Other particles in these theories could be dark matter if they were the lightest of the new particles (and satisfied cosmological and collider constraints), which depends on where exactly we sit in the rather large theoretical parameter space, but the neutralino and Kaluza-Klein (UED) photon are typically the lightest stable new particles.

\textbf{Axions:} Axions' ``buy one, get one free'' claim to fame is that they emerge out of a solution to the strong-CP problem in particle physics.  In the quantum chromodynamics (QCD) Lagrangian, there exists a term which allows significant but as-yet unobserved CP violation in QCD and contributes to the electric dipole moment of the neutron.  Upper limits on the neutron electric dipole moment suggest that the coefficient for this term should be $\lesssim 10^{-9}$, which smacks of fine tuning \cite{baker}.  Now, there is in principle nothing wrong with a parameter having a small value---on the neutrino side, the active neutrinos are at least six orders of magnitude smaller in mass than the next-lightest Standard Model particle, the electron \cite{deputter2012}.  But, usually when a parameter that could be huge is nearly zero, it implies that some sort of protective symmetry is at work.  The Peccei-Quinn solution to making this coefficient small is to turn that coefficient into a dynamical field, and add a global symmetry that, when broken, drives the offending term in the QCD Lagrangian to be precisely zero.  The new field's fluctuations about the new vacuum of the broken theory are axions, the pseudo-Nambu-Goldstone bosons of the broken symmetry.

Axions are in some ways less natural than WIMPs because it is tricky to get their comoving number density to match the observed dark-matter density.  There are a number of axion production mechanisms (all of which must be present to some extent), but the preferred way to produce dark-matter axions is through non-thermal coherent oscillations of the axion field near the QCD phase transition.  In that case, axions are light ($\sim 10 \mu$eV) and are born with no momentum.  See Chapter 10 of Ref. \cite{KT} for a review of axion production mechanisms.

\textbf{Gravitinos:}  While supersymmetric neutralinos are the dark-matter candidate of choice in some swaths of the MSSM, the gravitino, the supersymmetric partner of the graviton, may be dark matter in other swaths.  Depending on exactly how supersymmetry is broken, the gravitino could be anywhere in the mass range of $\sim$eV to TeV, although masses $\lesssim$keV are disfavored because they wash out too much small-scale structure (see Sec. \ref{sec:astro}; \cite{usefulparticle}).  In order for lighter gravitinos to be dark matter, one typically must introduce some non-standard cosmology \cite{fengsam}.  Heavy gravitinos are, in my opinion, more interesting.  If the next-lightest supersymmetric particle (NLSP) is only barely more massive than the gravitino, that particle species may be thermally produced and then decay at a later time to gravitinos.  Thus, even though gravitinos basically do not interact with the Standard Model (and thus would not typically be born as thermal relics), they can inherit the WIMP miracle from the NLSP.  The gravitino in this scenario is a ``superWIMP'' \cite{fengsuper}.  Because these massive gravitinos are born out of decays at relatively high momentum, they can smear out primordial density perturbations on small scales.  
Gravitinos are not nearly as beloved as WIMPs as dark-matter candidates because of the difficulty of getting the abundance just right and because they are much harder to detect using conventional methods.

\vskip 0.2cm
There are other dark-matter candidates that are plausible and solve some other problems in physics, although they to not provide quite the same bargain-hunting thrill of the previously discussed candidates.  I will list only two classes of candidate.

\textbf{Sterile neutrinos:} Sterile neutrinos are neutrinos that do not interact electroweakly.  Since mass eigenstates are not the same as the electroweak eigenstates (i.e., $\nu_e, \nu_\mu, \nu_\tau$), sterile neutrinos may mix with electroweak, or active, neutrinos.  Sterile neutrinos have been proposed in a number of contexts; they can be a mass-generating mechanism for the active neutrinos, they can simply be the right-handed counterparts to the active species, or explain certain neutrino-experiment anomalies \cite{sterile}.  As dark matter, sterile neutrinos may be created in the early Universe in a variety of ways.  Depending on their creation mechanism, they can be constrained by their effects on smaller-scale structure in the Universe \cite{lya_darkmatter}.  Because sterile neutrinos mix with active neutrinos, they have a small decay probability to an active neutrino and a photon \cite{sterilexray}.  The simplest model of sterile neutrino dark matter (Dodelson-Widrow neutrinos) are excluded by a combination of small-scale structure observations and non-detections of X-rays from galaxies \cite{lya_darkmatter,sterilexray} (for an alternative view, see \cite{kev}).

\textbf{Hidden-sector dark matter:} There is no reason to expect that the dark sector consists of only one or a handful of boring particles; after all, the Standard Model has richly interesting physics.  Extensions to the Standard Model open the door to other sectors of physics that may not have much contact with the Standard Model.  For example, supersymmetry has to be a broken theory, and the MSSM (the simplest supersymmetric extension to the Standard Model) is not going to break itself; new fields are needed to break supersymmetry and communicate that to the Standard Model.  Those fields may also communicate supersymmetry breaking to other sectors.  Sectors that have little communication with the Standard Model are called ``hidden'' or ``dark'' sectors.  A lot of interesting physics is allowed in the hidden sector, including the existence of ``dark photons'' \cite{hidden}.

\vskip 0.2cm
At the far other end of the spectrum, there are dark-matter candidates that are considered ``exotic'' or ``cooked up''.  These are typically highly specialized models designed to interpret so-called anomalies in cosmic-ray observations or particle-physics experiments as dark matter \cite{cranomaly}.  These candidates tend to have short lives but lead to interesting insights and new directions. 


Good reviews on particle dark-matter candidates are given in Refs. \cite{KT,usefulparticle}.  For an introduction to particle physics, I recommend Griffiths' book \cite{particle}.

\section{Astroparticle searches for dark matter}\label{sec:astroparticle}
\begin{figure}
\centering \includegraphics[width=0.4\textwidth,angle=90]{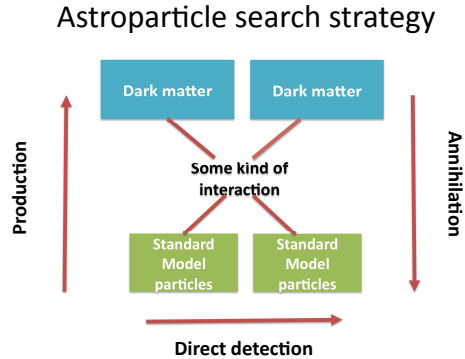}
\caption{\label{fig:astroparticle}Sketch of the different types of astroparticle search strategies for dark-matter detection.  The central figure is a toy Feynman diagram, and the search strategies depend on the direction in which one looks at the diagram.}
\end{figure}

Astroparticle searches depend on the type and strength of the interaction between dark matter and the Standard Model.  There are three main strategies for exploiting this interaction, as illustrated in Fig. \ref{fig:astroparticle}.  Looking at the diagram bottom-to-top, we produce dark-matter particles using Standard-Model particles.  This method is most commonly employed at large colliders (e.g., the Large Hadron Collider [LHC]) or using specialized experiments.  Reading Fig. \ref{fig:astroparticle} sideways, we look for the effects on Standard-Model particles induced by their interactions with dark-matter particles.  If we look at Fig. \ref{fig:astroparticle} top-to-bottom, we are lead to look for Standard-Model particles emerging from dark-matter annihilation or decays.

\subsection{WIMP searches}
Since WIMPs are the most popular class of dark-matter candidate (or at least the class which gets the most experiments), I will describe WIMP searches first and in the most detail.

\subsubsection{Colliders}
WIMPs will not directly be observed if they are created at colliders--given that they are neutral and weakly interacting, they are like gigantic neutrinos in terms of detection prospects.  However, it is possible to infer their existence.  The quarks and gluons in the protons smashed together at the LHC typically do not annihilate directly to WIMPs---since WIMPs belong to entire theories beyond the Standard Model, there are a panoply of other extra particles to which quarks and gluons may annihilate (e.g., colored particles like squarks and gluinos in the MSSM).  Those other particles may eventually decay to WIMPs inside the detector, the signature of which is missing energy when one tries to reconstruct the chain of events.  There has been a huge amount of effort to figure out which types of events (characterized by the number and types of jets, leptons, geometry, timing) are likely to lead to the best constraints on different WIMP models \cite{lhctheory}.  There is not yet experimental evidence of physics beyond the Standard Model \cite{lhcobs}.  Even if evidence for a WIMP is eventually found, we will not know if that particle is stable on timescales longer than a nanosecond.


\subsubsection{Direct detection}
Galactic WIMPs can ram into nuclei in the lab, depositing of order tens to hundreds of keV of kinetic energy to a single nucleus.  This is of order $10^7$ times less than the kinetic energy of a fruit fly, and the event rate is many orders of magnitude less than the ambient flux of cosmic rays, posing unique challenges to detection.  Nevertheless, there are dozens of experiments planned or underway to look for WIMPs this way \cite{ddstrategy}.  


The DAMA/LIBRA, CRESST, and CoGeNT experiments claim (sometimes in mild terms) WIMP detections \cite{ddclaim}.  It would be fair to say that these claims are not widely believed, especially given the null detections of other experiments.  Pretty much every experimentalist I have met has his or her own theory of the origin of the DAMA/LIBRA signal \cite{nygren}.  The DM-Ice collaboration is in the process of performing a DAMA-like experiment at the South Pole, ingeniously using the IceCube Neutrino Observatory as a cosmic-ray veto \cite{dmice}.  The best constraints from experiments that do not find significant events above background are XENON100, CDMS-II, and COUPP, and are cutting through swaths of WIMP model space \cite{ddresults}.  Currently, experiments are making rapid gains in sensitivity because it is possible (through great effort!) to do nearly zero-background searches, but soon (in the next decade) experiments will hit the wall of irreducible astrophysical neutrino backgrounds.  \emph{Neutrinos!}

\subsubsection{Annihilation}
The best places to look for WIMP annihilation are in dark-matter-dense objects, since the annihilation rate goes as the square of the density, and for which there are few other contaminating fore/backgrounds (or signals, depending on your point of view!).  Such objects include galaxy clusters, Milky Way dwarf galaxies, the Milky Way halo, the diffuse gamma-ray background (both the average signal and anisotropies), possible nearby dark-matter subhalos, and the center of the Sun \cite{cranomaly,ann_all,ann_dwarf,ann_sun}.  WIMPs annihilate to a wide variety of Standard-Model particles, but some of those particles are easier to search for than others.  There are some searches for WIMP annihilations to charged particles \cite{cranomaly}.  The two big problems with charged-particle searches are that even astrophysical emission mechanisms of charged particles are poorly understood and that charged particles have complicated diffusion histories, which are not nearly as well understood as it is sometimes made out to be.  I personally won't touch most charged-particle probes of dark matter with a 30-foot pole, but some people have done interesting work in this field.

Gamma rays and neutrinos point directly back to their sources, and are thus easier to interpret than charged particles.  Currently the most interesting constraints come from gamma-ray observations of the Milky Way halo and of the dwarf galaxies therein, and neutrino-telescope observations of the Sun \cite{ann_dwarf,ann_sun}.  The Milky Way dwarf galaxies are the most dark-matter-dense objects known, have few baryons, and are nearby, thus making them the perfect targets for WIMP-annihilation searches \cite{mwdwarfs}.  The gamma-ray flux limits from the Fermi Gamma-ray Space Telescope indicates that we are starting to cut through interesting WIMP parameter space.  The limits on gamma-ray annihilation in the Milky Way halo coming from Fermi and the ground-based H.E.S.S. telescope are only somewhat weaker and span a larger WIMP mass range than the current dwarf limits \cite{ann_dwarf}.

The Sun accumulates Galactic WIMPs when they scatter off solar nuclei to energies below the escape velocity of the Sun.  If the capture and annihilation rates of WIMPs in the Sun are in equilibrium, the annihilation rate is exactly half the capture rate, making solar WIMP searches sensitive to the elastic-scattering cross section.  Current constraints are competitive with direct-detection searches, even if there is still uncertainty in the capture-rate calculation \cite{ann_sun}.

\subsection{Other}\label{sec:particleaxion}
Dark-matter axions could be detected in laboratory experiments, exploiting the (quite weak) axions coupling to photons.  While axion-production and indirect-detection experiments do not yet probe cosmologically significant axion parameter space \cite{axionprod}, direct-detection searches will soon.  The ADMX experiment will be upgrading to Phase 2 this year, which has the potential to probe cosmologically-significant axions for two of the most popular QCD axion models \cite{admx}.  

The APEX experiment is searching for a light hidden-sector gauge boson that mixes with photons, currently reporting null results (although these are early days for the experiment)  \cite{apex}.

\section{The nature of dark matter through astronomical searches}\label{sec:astro}
However, if dark matter has only extremely weak couplings to the Standard Model, the astroparticle searches are dead on arrival.  
We will not necessarily be able to rule out candidates, merely rule out parts of their parameter space.  Thus, it would be great if we had some way of characterizing dark-matter physics that did not depend on Standard-Model interactions.  Fortunately, we have just such a thing!  Astronomical observations of the effects of the gravity of dark matter on baryons!  Recall that all we know about dark matter comes from exactly those ``gravitational probes'' of dark-matter physics (Sec. \ref{sec:intro} and \cite{skp}).

\subsection{Mapping dark-sector physics to observables}
In order to use astronomical observations to constrain dark-matter physics, we need to find a mapping between the two.  It is more useful to consider general dark-matter phenomenology than specific dark-matter models, at least at the present.  One way to classify dark-matter phenomenology is by physics important at early or late times.  This means of dark-matter classification is defined and explored in Ref. \cite{skp}.  

In the early Universe, the physics that matters most is the velocity distribution function of dark matter at its birth or freeze-out epoch.  Dark matter that freezes out or is created non-relativistic is called cold dark matter (CDM).  WIMPs and non-thermally-produced axions are CDM.  Inflation lays down density fluctuations (more precisely: fluctuations in the gravitational potential) on an incredibly wide range of scales, and the non-relativistic nature of CDM means that these fluctuations are left largely intact except on tiny scales related to the free-streaming length.  Hot dark matter (HDM) is dark matter that is born highly relativistic.  Because of its high speed, HDM can escape and thus wash out density perturbations on large scales in the early Universe.  HDM is constrained to make up a tiny percentage of the mass-energy density of the Universe \cite{deputter2012}.  In between these two extremes is warm dark matter (WDM).  Examples of WDM include gravitinos and sterile neutrinos.  We should see evidence for the temperature of dark matter at all observable epochs in the Universe.

The other dimension to dark-matter classification is its late-time behavior.  The dark-matter phenomenology that is important at late times is its stability to decays and self-interactions involving a hidden sector.  Self-interactions are more important at late times than early times because the self-interaction rate scales as the square of the dark-matter density.  There are simply more places in the Universe with high density at late times than early times.  Late-time effects can be distinguished from early-time effects because of the arrow of time.

The stable CDM paradigm is \emph{the top dog} among astrophysicists; nearly all structure-formation predictions are really stable CDM predictions (see \cite{cdmtheory} for reviews and references).  From simulations, we know how CDM structure evolves (at least in the absence of baryons) and how dark-matter halos cluster.  We find that dark-matter halos have cuspy density profiles, that halos are triaxial, and that the central density of halos depends on the mass of the halo.  Dark-matter halos have subhalo mass functions that extend down beyond the smallest simulated scales.

Stable WDM looks like stable CDM on scales $\gtrsim 10$ Mpc, but deviates below those scales as the speediness of WDM particles in the early Universe creates a cutoff in the matter power spectrum \cite{wdmlss}.  At late times, the evolution of the matter power spectrum is more subtle as halos form. Large dark-matter halos are virtually indistinguishable from stable CDM halos except that they may be somewhat less concentrated, but smaller halos, which form out of density perturbations near the cutoff scale in the power spectrum, look fluffier and less cuspy than CDM halos.  The subhalo mass function drops significantly on mass scales corresponding to that cutoff scale.

Unstable CDM deviates from stable CDM on large scales as well as small \cite{decaylss}.  If unstable CDM decays to relativistic particles, it changes the background evolution of the Universe.  Even if unstable CDM decays to non-relativistic particles, the particles stream out of dark-matter halos, causing the growth function of structure to acquire a scale dependence.  On smaller scales, halos are less dense than stable CDM halos due to the injection of kinetic energy into the halos from the decays.  The properties of subhalos have not been studied in great detail yet.  

Stable self-interacting CDM has made a bit of a theoretical comeback of late as part of the hidden-sector paradigm \cite{sidmth,hidden}.  Self-interacting CDM looks like stable CDM on large scales through cosmic time.  One finds deviations from CDM predictions only in the inner parts of dark-matter halos at late times.  The inner parts of halos to become cored and more spherical because of the exchange of energy and angular momentum among particles.  It is hypothesized that there will be a deficit of subhalos in the central regions of halos, but this prediction remains poorly quantified.


\subsection{Observations}
Currently, observations of large-scale structure (scales $\gtrsim 10$ Mpc) across cosmic time are consistent with a stable, cold-dark-matter picture \cite{lss,cmb,lya_darkmatter}.  Since self-interactions and WDM only show deviations from stable CDM on small scales, this implies that the observations on large scales are also consistent with the self-interacting CDM and WDM pictures.  Large-scale structure observations indicate that the lifetime of the parent dark-matter particle must be $\gtrsim 3$ times the Hubble time for recoil speeds of the daughter dark-matter particle of $\gtrsim 100\hbox{ km s}^{-1}$ \cite{decaylss}.  Most of the constraints on decaying dark matter emerge from the Sloan Digital Sky Survey and X-ray cluster counts.  Future large galaxy surveys, especially ones designed with dark-energy constraints in mind, will also constrain dark-matter models \cite{decaylss,desurvey}.  Next-generation galaxy surveys will probe large redshifts, thus allowing for tomographic studies of the late-time physics of dark matter.

Observations of small-scale structure (i.e., on scales of individual dark-matter halos) have the potential to be quite constraining, although in practice such observations are often difficult to interpret.  Observations of galaxy clusters and individual galaxies using strong lensing or galaxy-galaxy weak lensing indicate that dark-matter halos are indeed ellipsoidal, although a quantitative comparison with theoretical expectations is tricky \cite{clusters}.  There are hints from the smallest observed dark-matter halos (the halos of Milky Way dwarf galaxies) to the largest (galaxy clusters) that the density profiles are not well described by those found in stable CDM simulations without baryons \cite{halodensity}.  It is not clear yet if those deviations are a result of baryonic or dark-matter physics.

The subhalo mass function and subhalo central densities ought to be interesting probes of dark-matter physics \cite{wdmsubs}.  For (sub)halo masses smaller than $\sim 10^{10}M_\odot$, there are really only two ways to probe their mass function and central densities.  First, next-generation deep galaxy surveys should reveal more dwarf companions of the Milky Way, which may be characterized using existing techniques \cite{mwdwarfs,desurvey}.  However, this method relies on the existence of a decent number of stars in small dark-matter subhalos.  We do not really know how star formation proceeds in small halos.  It is better to not have to depend on baryons to probe such small halos.  Fortunately, we may look for subhalos using gravitational lensing.  In strong lenses, subhalos in the lens can change the positions and magnifications of the images, and perturb the light travel times \cite{vegetti,lenssub}.  I am part of the science team for the Observatory for Multi-Epoch Gravitational Lens Astrophysics (OMEGA) Explorer mission concept to monitor multiply-lensed active galactic nuclei for magnification an light arrival-time anomalies associated with subhalos in the lens galaxy \cite{omega}.  This is a unique way of probing dark-matter physics, and highly complementary to other dark-matter searches.  NASA should \emph{definitely} fund us in the next Explorer-class mission call!

\subsection{Caveats}\label{sec:astrocaveat}
Any interpretation of observations in the context of dark matter depends on a careful and accurate mapping of dark-matter physics to astronomical observables.  A HUGE source of systematics for this mapping is our ignorance of the specific ways in which galaxy evolution alters dark-matter halos and measures of the matter power spectrum \cite{hydrosim}.  Most of the predictions discussed in this section were made using dark-matter-only simulations.  However, we do not know the relative importance of various processes in galaxy evolution for dark-matter-halo evolution \cite{hydrores}.  Even when a subset of the physics we think must be important for galaxy evolution is included in simulations, the effects on dark-matter halos is extremely sensitive to the implementation of the galaxy physics in the codes \cite{hydrosim}.  One thing that appears to be important both for getting the dark-matter-halo morphologies as well as galaxy properties right is to resolve giant-molecular-cloud-sized regions \cite{hydrores}.  This is somewhat depressing because it is currently only possible to resolve such small scales for individual dwarf galaxies.  On the other hand, it implies job security for computational physicists.

\section{Conclusion}\label{sec:conclusion}
Neither astroparticle nor astronomical searches for dark matter are going to characterize dark matter on their own.  For example, say that in the next five years we find some sort of new, massive, neutral particle at the LHC, but do not see anything in direct-detection experiments or in neutrino telescopes or gamma-ray telescopes.  Is this new particle stable, and can it be all the dark matter?

Astronomical observations can answer these questions, or at least provide some guidance.  If the next generation of giant galaxy surveys sees some evidence of an anomalous scale-dependent growth of structure, it could hint that the dark matter is indeed unstable but with a long lifetime.  Thus the conventional WIMP model might be dead, but variants thereof may be alive.  On the other hand, if the largest scales of the Universe evolve as they would for stable CDM but dark-matter halos continue to look somewhat cored, and if OMEGA finds a suppressed subhalo mass function in lens galaxies, then this might indicate a significant amount of self interaction in a hidden-sector model.  Or it is possible that there is no deviation from stable CDM predictions, and we conclude that even if the particle found at the LHC is not all of the dark matter, dark matter must be pretty stable, fairly weakly interacting, and cold.

In the next decade, I think it will be important to recognize our prejudices and confront them head on, namely that we as a community are thoroughly captivated by stable CDM WIMP dark matter.  But just because WIMPs are beautiful dark-matter candidates does not mean that dark matter \emph{must} consist of WIMPs.  In this spirit, I will close with a pair of quotes from famous scientists.  From Steven Weinberg, ``It seems that scientists are often attracted to beautiful theories in the way that insects are attracted to flowers---not by logical deduction, but by something like a sense of smell.''  And from Carl Sagan, ``With insufficient data it is easy to go wrong.''  We all hope that we will soon be in an era of abundant data.  The key will be to see how all these different searches fit together to present a unified picture of the nature of dark matter.


\begin{thebibliography}{99}

  \bibitem{lss} B. Reid et al., \emph{Cosmological constraints from the clustering of the Sloan Digital Sky Survey DR7 luminous red galaxies}, MNRAS \textbf{404}, 60 (2010); W. Percival et al., \emph{Baryon acoustic oscillations in the Sloan Digital Sky Survey Data Release 7 galaxy sample}, MNRAS \textbf{401}, 2148 (2010); E.~M.~Huff et al., \emph{Seeing in the dark II: Cosmic shear in the Sloan Digital Sky Survey}, arXiv:1112.3143, J.~L.~Tinker et al., \emph{Cosmological Constraints from Galaxy Clustering and the Mass-to-number Ratio of Galaxy Clusters}, ApJ \textbf{745}, 16 (2012).

  \bibitem{cmb} D.~Larson et al., \emph{Seven-Year Wilkinson Microwave Anisotropy Probe (WMAP) Observations: Power Spectra and WMAP-Derived Parameters}, ApJS \textbf{192}, 16 (2011); J.~Dunkley et al., \emph{The Atacama Cosmology Telescope: Cosmological Parameters from the 2008 Power Spectrum}, ApJ \textbf{739}, 52 (2011)

  \bibitem{cmbtutorial} Wayne Hu maintains excellent tutorials on the physics of the cosmic microwave background and large-scale structure.  The interested reader is enthusiastically recommended to check out http://background.uchicago.edu/index.html


  \bibitem{steigman2007} R.~H.~Cyburt, \emph{Primordial nucleosynthesis for the new cosmology: Determining uncertainties and examining concordance}, Phys. Rev. D \textbf{70}, 023505 (2004); G.~Steigman, \emph{Primordial Nucleosynthesis in the Precision Cosmology Era}, Annual Review of Nuclear and  Particle Science \textbf{57}, 463 (2007); F.~Iocco et al., \emph{Primordial nucleosynthesis: From precision cosmology to fundamental physics}, Phys. Rep. \textbf{472}, 1 (2009); B.~Fields, \emph{The Primordial Lithium Problem}, Annual Review of Nuclear and Particle Science \textbf{61}, 47 (2011).

  \bibitem{axions} R.~D.~Peccei \& H.~R.~Quinn, \emph{Constraints imposed by CP conservation in the presence of pseudoparticles}, Phys. Rev. D \textbf{16}, 1791 (1977); S.~Weinberg, \emph{A new light boson?}, Phys. Rev. Lett. \textbf{40}, 223 (1978); F.~Wilczek, \emph{Problem of strong P and T invariance in the presence of instantons}, Phys. Rev. Lett. \textbf{40}, 279 (1978).

  \bibitem{lya_darkmatter} K.~Abazajian, \emph{Linear cosmological structure limits on warm dark matter}, Phys. Rev. D \textbf{73}, 063513 (2006); M.~Viel et al., \emph{Can Sterile Neutrinos Be Ruled Out as Warm Dark Matter Candidates?}, Phys. Rev. D \textbf{97}, 071301 (2006); A.~Boyarsky et al., \emph{Lyman-$\alpha$ constraints on warm and on warm-plus-cold dark matter models}, JCAP \textbf{005} (2009) 012.

  \bibitem{em} K.~Sigurdson et al., \emph{Dark-matter electric and magnetic dipole moments}, Phys. Rev. D \textbf{70}, 083501 (2004); S.~D.~McDermott, H.-B.~Yu, \& K.~M.~Zurek, \emph{Turning off the lights: How dark is dark matter?}, Phys. Rev. D \textbf{83}, 063509 (2011)

  \bibitem{sidmth} D.~N.~Spergel \& P.~J.~Steinhardt, \emph{Observational Evidence for Self-Interacting Cold Dark Matter}, Phys. Rev. Lett. \textbf{84}, 3760 (2000); M.~R.~Buckley \& P.~J.~Fox, \emph{Dark matter self-interactions and light force carriers}, Phys. Rev. D \textbf{81}, 083522 (2010); J.~L.~Feng, M.~Kaplinghat, \& H.-B.~Yu, \emph{Halo-Shape and Relic-Density Exclusions of Sommerfeld-Enhanced Dark Matter Explanations of Cosmic Ray Excesses}, Phys. Rev. Lett. \textbf{104}, 151301 (2010).

  \bibitem{sidmobs} O.~Y.~Gnedin \& J.~P.~Ostriker, \emph{Limits on Collisional Dark Matter from Elliptical Galaxies in Clusters}, ApJ \textbf{561}, 61 (2001); J.~F.~Hennawi \& J.~P.~Ostriker, \emph{Observational Constraints on the Self-interacting Dark Matter Scenario and the Growth of Supermassive Black Holes}, ApJ \textbf{572}, 41 (2002); J.~Miralda-Escud{\'e}, \emph{A Test of the Collisional Dark Matter Hypothesis from Cluster Lensing}, ApJ \textbf{564}, 60 (2002); S.~W.~Randall et al., \emph{Constraints of the Self-Interaction Cross Section of Dark Matter from Numerical Simulations of the Merging Galaxy Cluster 1E 0657-56}, ApJ \textbf{679}, 1173 (2008).

  \bibitem{vegetti} S.~Vegetti et al., \emph{Detection of a dark substructure through gravitational imaging}, MNRAS \textbf{408}, 1969 (2010).

  \bibitem{mwdwarfs} L.~E.~Strigari et al., \emph{A common mass scale for satellite galaxies of the Milky Way}, Nature \textbf{454}, 1096 (2008); M.~G.~Walker et al., \emph{A Universal Mass Profile for Dwarf Spheroidal Galaxies?}, ApJ \textbf{704}, 1274 (2009); J.~Wolf et al., \emph{Accurate masses for dispersion-supported galaxies}, MNRAS \textbf{406}, 1220 (2010).

  \bibitem{pbh} B.~J.~Carr et al., \emph{Cosmological constraints on primordial black holes}, Phys. Rev. D \textbf{81}, 104019 (2010).

  \bibitem{steigman1985} G.~Steigman \& M.~S.~Turner, \emph{Cosmological constraints on the properties of weakly interacting massive particles}, Nucl. Phys. B \textbf{253}, 375 (1985).

  \bibitem{astronomerunit} For readers unfamiliar to particle-physics units, you need to know the following: all physical quantities are related to energy.  Thus, masses are given in units of rest-mass energy, lengths are described in terms of the energy a photon with the relevant length-scale as the wavelength, and units of time are related to the energy of a photon with a period of $2\pi/t$.  For further information on unit conversion, see Appendix A of \cite{KT}.

  \bibitem{KT} E.~W.~Kolb \& M.~S.~Turner, \emph{The Early Universe}, Addison Wesley, New York 1990.


  \bibitem{griest} K.~Griest, \emph{Cross sections, relic abundance, and detection rates for neutralino dark matter}, Phys. Rev. D \textbf{38}, 2357 (1988); H.-C.~Cheng, J.~L.~Feng, \& K.~T.~Matchev, \emph{Kaluza-Klein Dark Matter}, Phys. Rev. Lett. \textbf{89}, 211301 (2002); G.~Servant \& T.~M.~P.~Tait, \emph{Is the lightest Kaluza-Klein particle a viable dark matter candidate?}, Nucl. Phys. B. \textbf{650}, 391 (2003).

  \bibitem{baker} C.~A.~Baker et al., \emph{Improved Experimental Limit on the Electric Dipole Moment of the Neutron}, Phys. Rev. Lett. \textbf{97}, 13801 (2006).

  \bibitem{deputter2012} K.~Abazajian et al., \emph{Nonlinear cosmological matter power spectrum with massive neutrinos: The halo model}, Phys. Rev. D \textbf{71}, 043507 (2005); R. de Putter et al., arXiv:1201.1909.

  \bibitem{usefulparticle} G.~Jungman, M.~Kamionkowski, \& K.~Griest, \emph{Supersymmetric dark matter}, Phys. Rep. \textbf{267}, 195 (1996); S.~J.~Asztalos et al., \emph{Searches for Astrophysical and Cosmological Axions}, Annual Review of Nuclear and Particle Systems \textbf{56}, 293 (2006); D.~Hooper \& S.~Profumo, \emph{Dark matter and collider phenomenology of universal extra dimensions}, Phys. Rep. \textbf{453}, 29 (2007); J.~L.~Feng, \emph{Dark Matter Candidates from Particle Physics and Methods of Detection}, ARA\&A \textbf{48}, 495 (2010).

  \bibitem{fengsam} 
J.~L.~Feng, M.~Kamionkowski, \& S.~K.~Lee, \emph{Light gravitinos at colliders and implications for cosmology}, Phys. Rev. D \textbf{82}, 015012 (2010).

  \bibitem{fengsuper} J.~L.~Feng, A. Rajaraman, \& F.~Takayama, \emph{Superweakly Interacting Massive Particles}, Phys. Rev. Lett. \textbf{91}, 011302 (2003).

  \bibitem{sterile} S.~Dodelson \& L.~M.~Widrow, \emph{Sterile neutrinos as dark matter}, Phys. Rev. Lett. \textbf{72}, 17 (1994); X.~Shi \& G.~M.~Fuller, \emph{New Dark Matter Candidate: Nonthermal Sterile Neutrinos}, Phys. Rev. Lett. \textbf{82}, 2832 (1999); K.~Petraki \& A.~Kusenko, \emph{Dark-matter sterile neutrinos in models with a gauge singlet in the Higgs sector}, Phys. Rev. D \textbf{77}, 065014 (2008); A.~Kusenko, \emph{Sterile neutrinos: The dark side of the light fermions}, Phys. Rep. \textbf{481}, 1 (2009); LSND, C. Athanassopoulos, et al., \emph{Results on $\nu_\mu\rightarrow \nu_e$ Neutrino Oscillations fron the LSND Experiment}, Phys. Rev. Lett. \textbf{81}, 1774 (1998).

  \bibitem{sterilexray} K.~Abazajian, G.~M.~Fuller, \& W.~H.~Tucker, \emph{Direct Detection of Warm Dark Matter in the X-Ray}, ApJ \textbf{562}, 593 (2001); C.~R.~Watson, Z.~Li, \& N.~K.~Polley, \emph{Constraining Sterile Neutrino Warm Dark Matter with Chandra Observations of the Andromeda Galaxy}, arXiv:1111.4217. 

  \bibitem{kev} K. Abazajian.
 
  \bibitem{hidden} J.~L.~Feng \& J.~Kumar, \emph{Dark-Matter Particles without Weak-Scale Masses or Weak Interactions}, Phys. Rev. Lett. \textbf{101}, 231301 (2008); L.~Ackerman et al., \emph{Dark matter and dark radiation}, Phys. Rev. D \textbf{79}, 023519 (2009); 
J.~L.~Feng et al., \emph{Hidden charged dark matter}, JCAP \textbf{07} (2009) 004; J.~L.~Feng, V.~Rentala, \& Z.~Surujon, \emph{WIMPless dark matter in anomaly-mediated supersymmetry breaking with hidden QED}, Phys. Rev. D \textbf{84}, 095033 (2011); M.~Pospelov, A.~Ritz, \& M.~Voloshin, \emph{Secluded WIMP dark matter}, Phys. Lett. B \textbf{662}, 53 (2008); N.~Arkani-Hamed et al., \emph{A theory of dark matter}, Phys. Rev. D \textbf{79}, 015014 (2009); D.~E.~Kaplan, M.~A.~Luty, \& K.~M.~Zurek, \emph{Asymmetric dark matter}, Phys. Rev. D \textbf{79}, 115016 (2009); T.~Cohen et al., \emph{Asymmetric dark matter from a GeV hidden sector}, Phys. Rev. D \textbf{82}, 056001 (2010).

  \bibitem{cranomaly} O.~Adriani et al., \emph{An anomalous positron abundance in cosmic rays with energies 1.5-100 GeV}, Nature \textbf{458}, 607 (2009); A.~A.~Abdo et al., \emph{Measurement of the Cosmic Ray $e^+ + e^-$ Spectrum from 20 GeV to 1 TeV with the Fermi Large Area Telescope}, Phys. Rev. Lett. \textbf{102}, 181101 (2009).

  \bibitem{particle} D.~Griffiths, \emph{Introduction to Elementary Particles}, Wiley, New York, 2008.

  \bibitem{lhctheory} J.~L.~Feng, J.-F.~Grivaz, \& J.~Nachtman, \emph{Searches for supersymmetry at high-energy colliders}, Rev. Mod. Phys. \textbf{82}, 699 (2010); D.~Alves et al., \emph{Simplified Models for LHC New Physics Searches}, arXiv:1105.2838.


  \bibitem{lhcobs} 
S. Chatrchyan et al., \emph{Search for supersymmetry in pp collisions at $\sqrt{s} = 7$ TeV in events with a single lepton, jets, and missing transverse momentum}, JHEP \textbf{08} (2001) 156; ATLAS Collaboration, \emph{ Search for supersymmetry in pp collisions at s=7 TeV in final states with missing transverse momentum and b-jets}, Phys. Lett. B \textbf{701}, 398 (2011).



  \bibitem{ddstrategy} See the Dark Matter Working Group DUSEL White paper (2010) for a comprehensive list and review of direct-detection experiments, and references to those experiments: http://dmtools.brown.edu/DMWiki/index.php/File:DMWG\_DUSEL\_White\_Paper\_April\_2010.pdf

  \bibitem{ddclaim} R.~Bernabei et al., \emph{New results from DAMA/LIBRA}, Eur. Phys. J. C \textbf{67}, 39 (2010); C.~E.~Aalesth et al., \emph{Results from a Search for Light-Mass Dark Matter with a p-Type Point Contact Detector}, Phys. Rev. Lett. \textbf{106}, 131301 (2011); G.~Angloher et al., \emph{Results from 730 kg days of the CRESST-II Dark Matter Search}, arXiv:1109.0702.

  \bibitem{nygren} One popular theory for the origin of the DAMA/LIBRA signal is introduced in D.~Nygren, \emph{A testable conventional hypothesis for the DAMA-LIBRA annual modulation}, arXiv:1102.0815.

  \bibitem{dmice} J.~Cherwinka et al., \emph{A Search for the Dark Matter Annual Modulation in South Pole Ice}, arXiv:1106.1156.

  \bibitem{ddresults} E.~Behnke et al. \emph{Improved Limits on Spin-Dependent WIMP-Proton Interactions from a Two Liter CF$_3$I Bubble Chamber}, Phys. Rev. Lett. \textbf{106}, 021303 (2011); E.~Aprile et al., \emph{Dark Matter Results from 100 Live Days of XENON100 Data}, Phys. Rev. Lett. \textbf{107}, 13102 (2011); Z.~Ahmed, \emph{Combined limits on WIMPs from the CDMS and EDELWEISS experiments}, Phys. Rev. D \textbf{84}, 011102 (2011).


  \bibitem{ann_all} A.~A.~Abdo et al., \emph{Constraints on cosmological dark matter annihilation from the Fermi-LAT isotropic diffuse gamma-ray measurement}, JCAP \textbf{04} (2010) 014; R.~Abbasi et al., \emph{Search for dark matter from the Galactic halo with the IceCube Neutrino Telescope}, Phys. Rev. D \textbf{84}, 022004 (2011); A.~Pinzke, C.~Pfrommer, \& L.~Bergstr{\"o}m, \emph{Prospects of detecting gamma-ray emission from galaxy clusters: Cosmic rays and dark matter annihilations}, Phys. Rev. D \textbf{84}, 123509 (2011); A.~Cuoco et al., \emph{Anisotropies in the diffuse gamma-ray background measured by Fermi LAT}, arXiv:1110.1047.

  \bibitem{ann_dwarf} K.~N.~Abazajian et al., \emph{Conservative constraints on dark matter from the Fermi-LAT isotropic diffuse gamma-ray background spectrum}, JCAP \textbf{11} (2010) 041;  A.~Abramowski et al., \emph{Search for a Dark Matter Annihilation Signal from the Galactic Center Halo with H.E.S.S.}, Phys. Rev. Lett \textbf{106}, 161301 (2011); A.~Geringer-Sameth \& S.~M.~Koushiappas, \emph{Exclusion of canonical WIMPs by the joint analysis of Milky Way dwarfs with Fermi}, Phys. Rev. Lett. \textbf{107}, 241303 (2011); The Fermi LAT Collaboration, M.~Kaplinghat, \& G.~D.~Martinez, \emph{Constraining Dark Matter Models from a Combined Analysis of Milky Way Satellites with the Fermi Large Area Telescope}, Phys. Rev. Lett. \textbf{107}, 241302 (2011); R.~C.~Cotta et al., \emph{Constraints on the pMSSM from LAT Observations of Dwarf Spheroidal Galaxies}, arXiv:1111.2604.

  \bibitem{ann_sun} A.~Gould, \emph{Weakly interacting massive particle distribution in and evaporation from the sun}, ApJ \textbf{321}, 560 (1987); A.~H.~G.~Peter, \emph{Dark matter in the Solar System II. WIMP annihilation rates in the Sun}, Phys. Rev. D \textbf{79}, 103532 (2009); T. Tanaka et al., \emph{An Indirect Search for Weakly Interacting Massive Particles in the Sun Using 3109.6 Days of Upward-going Muons in Super-Kamiokande}, ApJ \textbf{742}, 78 (2011); The IceCube Collaboration, \emph{Multi-year search for dark matter annihilations in the Sun with the AMANDA-II and IceCube detectors}, arXiv:1112.1840; S.~Sivertsson \& J. Edsj{\"o}, \emph{WIMP diffusion in the solar system including solar WIMP-nucleon scattering}, arXiv:1201.1895.

  \bibitem{axionprod} K.~van Bibber et al., \emph{Proposed experiment to produce and detect light pseudoscalars}, Phys. Rev. Lett. \textbf{59}, 759 (1987); A.~S.~Chou et al., \emph{Search for Axionlike Particles Using a Variable-Baseline Photon-Regeneration Technique}, Phys. Rev. Lett. \textbf{100}, 080402 (2008); http://cast.web.cern.ch/CAST/

  \bibitem{admx} P.~Sikivie, \emph{Experimental tests of the 'invisible' axion}, Phys. Rev. Lett. \textbf{51}, 1415 (1983); http://www.phys.washington.edu/groups/admx/home.html


  \bibitem{apex} S.~Abrahamyan et al., \emph{Search for a New Gauge Boson in Electron-Nucleus Fixed-Target Scattering by the APEX Experiment}, Phys. Rev. Lett. \textbf{107}, 191804 (2011).


  \bibitem{skp} K.~Sigurdson, M.~Kaplinghat, \& A.~H.~G.~Peter, \emph{Gravitational probes of dark-matter physics}, in prep.


  \bibitem{cdmtheory} P.~J.~E.~Peebles, \emph{The large-scale structure of the universe}, Princeton University, Princeton, NJ, 1980; S.~Dodelson, \emph{Modern Cosmology}, Academic Press, Amsterdam, 2003; H.~Mo, F.~C.~van den Bosch, \& S.~White, \emph{Galaxy Formation and Evolution}, Cambridge University, Cambridge, UK, 2010.


  \bibitem{wdmlss} 
P.~Col{\'{\i}}n, O.~Valenzuela, \& V.~Avila-Reese, \emph{On the Structure of Dark Matter Halos at the Damping Scale of the Power Spectrum with and without Relict Velocities}, ApJ \textbf{673}, 203 (2008); F.~Villaescusa-Navarro \& N.~Dalal, \emph{Cores and cusps in warm dark matter halos}, JCAP \textbf{03} (2011) 024; M.~Viel et al., \emph{The Non-Linear Matter Power Spectrum in Warm Dark Matter Cosmologies}, arXiv:1107.4094; R.~M.~Dunstan et al., \emph{The Halo Model of Large Scale Structure for Warm Dark Matter}, arXiv:1109.6291; R.~E.~Smith \& K.~Markovic, \emph{Testing the warm dark matter paradigm with large-scale structures}, Phys. Rev. D \textbf{84}, 063507 (2011).

  \bibitem{decaylss} A.~H.~G.~Peter, \emph{Mapping the allowed parameter space for decaying dark matter models}, Phys. Rev. D \textbf{81}, 083511 (2010); A.~H.~G.~Peter, C.~E.~Moody, \& M.~Kamionkowsi, \emph{Dark-matter decays and self-gravitating halos}, Phys. Rev. D \textbf{81}, 103501 (2010); M.-Y.~Wang \& A.~R.~Zentner, \emph{Weak gravitational lensing as a method to constrain unstable dark matter}, Phys. Rev. D \textbf{82}, 123507 (2010); M.-Y.~Wang \& A.~R.~Zentner, \emph{Effects of Unstable Dark Matter on Large-Scale Structure and Constraints from Future Surveys}, arXiv:1201.2426.

  \bibitem{desurvey} PanSTARRS: http://pan-starrs.ifa.hawaii.edu/public/; SkyMapper: http://msowww.anu.edu.au/skymapper/; Dark Energy Survey: http://www.darkenergysurvey.org/; LSST: http://www.lsst.org/lsst/; WFIRST: http://wfirst.gsfc.nasa.gov/; BOSS: http://cosmology.lbl.gov/BOSS/; Prime Focus Spectrograph: http://sumire.ipmu.jp/en/

  \bibitem{clusters} R.~Mandelbaum et al., \emph{Density profiles of galaxy groups and clusters from SDSS galaxy-galaxy weak lensing}, MNRAS \textbf{372}, 758 (2006); R.~Mandelbaum et al., \emph{Ellipticity of dark matter haloes with galaxy-galaxy weak lensing}, MNRAS \textbf{370}, 1008 (2006); D.~J.~Sand et al., \emph{Separating Baryons and Dark Matter in Cluster Cores: A Full Two-dimensional Lensing and Dynamic Analysis of Abell 383 and MS 2137-23}, ApJ \textbf{674}, 711 (2008); J.~Richard et al., \emph{LoCuSS: First Results from Strong-lensing Analysis of 20 Massive Galaxy Clusters at $z\sim0.2$}, MNRAS \textbf{404}, 325 (2010).

  \bibitem{halodensity} R.~Kuzio de Naray, S.~S.~McGaugh, \& W.~J.~G.~de Blok, \emph{Mass Models for Low Surface Brightness Galaxies with High-Resolution Optical Velocity Fields}, ApJ \textbf{676}, 920 (2008); R.~Mandelbaum, U.~Seljak, \& C.~M.~Hirata, \emph{A halo mass---concentration relation from weak lensing}, JCAP \textbf{08} (2008) 006; M.~Boylan-Kolchin, J.~S.~Bullock, \& M.~Kaplinghat, \emph{Too big to fail? The puzzling darkness of massive Milky Way subhaloes}, MNRAS \textbf{415}, L40 (2011).

  \bibitem{wdmsubs} A.~H.~G.~Peter \& A.~J.~Benson, \emph{Dark-matter decays and Milky Way satellite galaxies}, Phys. Rev. D \textbf{82}, 123521 (2010); E.~Polisensky \& M.~Ricotti, \emph{Constraints on the dark matter particle mass from the number of Milky Way satellites}, Phys. Rev. D \textbf{83}, 043506 (2011), M.~Lovell et al., \emph{The Haloes of Bright Satellite Galaxies in a Warm Dark Matter Universe}, arXiv:1104.2929.

  \bibitem{lenssub} N.~Dalal \& C.~S.~Kochanek, \emph{Direct Detection of Cold Dark Matter Substructure}, ApJ \textbf{572}, 25 (2002); C.~R.~Keeton \& L.~A.~Moustakas, \emph{A New Channel for Detecting Dark Matter Substructure in Galaxies: Gravitational Lens Time Delays}, ApJ \textbf{699}, 1720 (2009); P.~J.~Marshall et al., \emph{Dark Matter Structures in the Universe: Prospects for Optical Astronomy in the Next Decade}, Astro2010 Decadal Survey (2010).

  \bibitem{omega} L.~Moustakas et al., \emph{The Observatory for Multi-Epoch Gravitational Lens Astrophysics (OMEGA)}, SPIE Conference Series \textbf{7010}, 41 (2008).

  \bibitem{hydrosim} D.~H.~Rudd, A.~R.~Zentner, \& A.~V.~Kravtsov, \emph{Effects of Baryons and Dissipation on the Matter Power Spectrum}, ApJ \textbf{672}, 19 (2008); S.~Pedrosa, P.~B.~Tissera, \& C.~Scannapieco, \emph{The joint evolution of baryons and dark matter halos}, MNRAS \textbf{402}, 776 (2010); C.~Scannapieco et al., \emph{The Aquila Comparison Project: The Effects of Feedback and Numerical Methods on Simulations of Galaxy Formation}, arXiv:1112.0315.


  \bibitem{hydrores} F.~Governato et al., \emph{Bulgeless dwarf galaxies and dark matter cores from supernova-driven outflows}, Nature \textbf{463}, 203 (2010); P.~F.~Hopkins, E.~Quataert, \& N.~Murray, \emph{Self-regulated star formation in galaxies via momentum input from massive stars}, MNRAS \textbf{417}, 950 (2011); M.~Kuhlen et al., \emph{Dwarf galaxy formation with H2-regulated star formation}, arXiv:1105.2376.


\end{thebibliography}
\end{document}